\documentclass[aps,11pt,prc,preprint,superscriptaddress,nofootinbib]{revtex4}
\usepackage[usenames]{color}
\usepackage{graphicx}
\usepackage{amsmath}
\usepackage{amsfonts}
\usepackage{amssymb}
\usepackage{mathrsfs}
\usepackage{bm}
\usepackage{verbatim}

\usepackage{ulem}

\newcommand{\beq}{\begin{equation}}
\newcommand{\eeq}{\end{equation}}
\newcommand{\bea}{\begin{eqnarray}}
\newcommand{\eea}{\end{eqnarray}}

\graphicspath{{../}}

\begin{document}

\title{The link between two-body model space and many-body model space}
\author{C. J. Yang}
\affiliation{Institut de Physique Nucl\'eaire, IN2P3-CNRS, Universit\'e Paris-Sud,
Universit\'e Paris-Saclay, F-91406 Orsay Cedex, France}
\affiliation{Institute for Nuclear Studies, Department of Physics, George Washington
University, Washington DC 20052, USA}
\email{yangjerry@gwu.edu}
\date{\today }

\begin{abstract}
An exact relation which links the ideal model space to be used in A-body
calculations when the two-body interaction is given in a truncated model
space is derived. Its implications on the effective field theory (EFT)
approach to no-core-shell-model (NCSM)\cite{Ia07a} is analyzed. Some
insights regarding whether details of two-body interaction becomes less
important in the calculations of many-body system are given. The result
suggests that there might be a way to establish an EFT expansion for heavy
nuclei and nuclear matter with an effective intereaction which has a much
simpler form than the nucleon-nucleon (NN) interaction in the vacuum. 
\end{abstract}

\pacs{67.85.Lm,21.65.-f}
\keywords{Effective field theory, nuclear structure, nuclear matter}
\maketitle


\section{Introduction}

Due to the limitation of computational power and limited knowledge about the
nature, calculations of physical systems always needed to be performed in
certain model space. In ab-initio calculations, when the inter-particle
interaction is defined through a transformation from the infinite space\footnote{%
This includes interactions which are regularized through regulators but only
vanishing exactly to zero at infinite momentum and those have been tamed by
unitary transformations such as similarity renormalization group (SRG)\cite%
{srg}.}, one usually needs to increase the model space in the many-body
calculations until a pattern of convergence is observed or can be
extrapolated. Various extrapolation methods, especially the extrapolation
regarding infrared cutoffs, have been studied recently\cite{extra}. On the
other hand, one could define the interaction directly in a given finite
model space, effects generated by physics out of the model space can be
included systematically through effective field theory (EFT), and are
represented by the low energy constants (LECs) associated with counter terms
after renormalization. In nuclear structure calculations, this direction has
been advocated by the Arizona group \cite{Ia07a,Ia07b,Ia10,Ji10,Ia10a,RJ12}
and carried out recently in Refs.\cite{oak,hobet,hobet2,yang16}. In these
approaches, since the interactions are built in a given finite model space,
one expects that the results of many-body calculations become exact also in
a finite model space. One then needs to answer the following question:
Suppose the interaction between two particles\footnote{%
Here the particles are just the basic degree of freedom considered in an
EFT, and they need not to be elementary particles.} is built and
renormalized within their center of mass (c.m.) momentum $|\mathbf{p}|\in
\lbrack 0,\Lambda ]$ and disappears elsewhere, then, for a system consists
of A particles through the same interaction, to what maximum momentum (with
respect to the c.m. of the system) will the system still response to an
outer probe? This maximum momentum is linked to the ideal model space
to be used in any A-body calculation. In the following we derive an exact
formula which links the two- and A-body model space through an analysis in
Jacobi coordinate.

\section{Derivation}

\bigskip We start with the usual definition of Jacobi momentums:%
\begin{eqnarray}
\vec{\pi _{1}} &=&\frac{1}{2}(\vec{p_{1}}-\vec{p_{2}}),  \label{j1} \\
\vec{\pi _{2}} &=&\frac{2}{3}\left( \vec{p_{3}}-\frac{1}{2}(\vec{p_{1}}+\vec{%
p_{2}})\right) ,  \label{j2} \\
\vec{\pi _{3}} &=&\frac{3}{4}\left( \vec{p_{4}}-\frac{1}{3}(\vec{p_{1}}+\vec{%
p_{2}}+\vec{p_{3}})\right) ,  \label{j3} \\
&:&  \notag \\
\vec{\pi }_{N-1} &=&\frac{N-1}{N}\left( \vec{p_{N}}-\frac{1}{N-1}(\vec{p_{1}}%
+\vec{p_{2}}+...+\vec{p}_{N-1})\right) .  \label{jaco}
\end{eqnarray}

Here $\vec{p_{i}}$ denotes the momentum of particle $i$ in the reference
frame, and the Jacobi momentum $\vec{\pi}_{N-1}$ denotes the relative
momentum of particle N with respect to the sub-system with one less
particle. Denote $\vec{\Lambda }_{ij}=|\vec{\Lambda }_{ij}| \widehat{n}_{ij}=$ $\frac{1}{%
2}(\vec{p}_{i}-\vec{p}_{j})=\frac{1}{2}\vec{p}_{ij}$, where $|\vec{\Lambda }_{ij}|\in
\lbrack 0,\Lambda ]$, with $\Lambda $ the
two-body cutoff, which defines the maximum model space the two body system
lives, and $\widehat{n}_{ij}$ the unit vector pointing from particle $i$ to $%
j$, then from Eq. (\ref{j1}) and (\ref{j2}) one obtains:%
\begin{eqnarray}
2\vec{\pi} _{1}+3\vec{\pi}_{2} &=&\vec{p}_{1}-\vec{p}_{2}+2\vec{p}_{3}-(%
\vec{p}_{1}+\vec{p}_{2}),  \notag \\
&=&4\cdot \frac{1}{2}(\vec{p}_{3}-\vec{p}_{2})=4\vec{\Lambda}_{32}.
\label{e3}
\end{eqnarray}%
Using $\vec{\pi} _{1}=\frac{1}{2}(\vec{p}_{1}-\vec{p}_{2})=$ $\vec{\Lambda}%
_{12}$, one obtains%
\begin{eqnarray}
2\vec{\Lambda}_{12}+3\vec{\pi} _{2} &=&4\vec{\Lambda}_{32},  \notag \\
3\vec{\pi}_{2} &=&4\vec{\Lambda}_{32}-2\vec{\Lambda}_{12}.
\end{eqnarray}%
Since there is no restriction on the alignment between $\vec{\Lambda}_{32}$ and $\vec{\Lambda}_{12}$, the maximum value $|\vec{\pi}_{2}|$ can have is $2\Lambda $.

\begin{figure}[h]
\includegraphics[width=7cm]{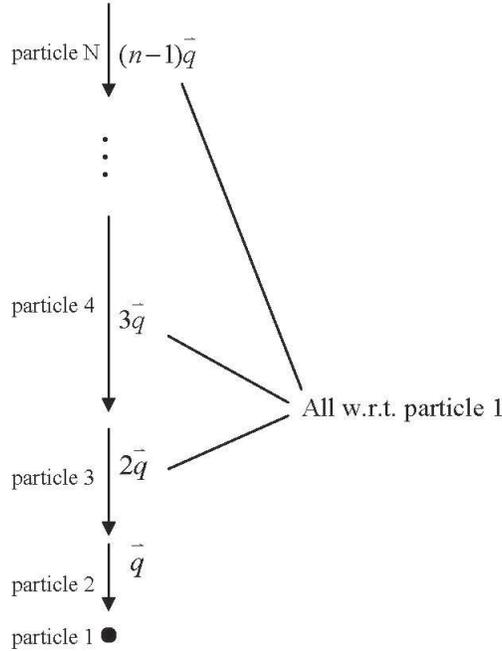}
\caption{A visual illustration of the extreme case where $N$ particles
interact within their maximum model space.}
\label{vis}
\end{figure}

The above maximum corresponds to the scenario that all 3 particles are
aligned and moving in the same direction, where particle 2 has the maximum
allowed relative momentum $|\vec{p}_{12}|_{\max}\equiv |\vec{p}_{1}-\vec{p}%
_{2}|_{\max}=2\Lambda $ with respect to particle 1, while particle 3 is moving
toward particle 1 with momentum $|\vec{p}_{13}|=4\Lambda $. In this case,
particle 3 will not interact with particle 1, but will still interact with
particle 2 since $|\vec{p}_{23}|=2\Lambda $ is within the maximum allowed
relative momentum between a pair. Actually, this extreme case can be
generalized to the N particles case as listed in Fig. \ref{vis}, where all
particles are aligned in the same direction with ascending increment of
relative momentum $\vec{q}$ $(|\vec{q}|=2\Lambda )$ respect to particle 1.
In this case, the $N^{th}$ particle only interacts with $(N-1)^{th}$
particle and nothing else, which guarantees that the largest possible model
space has been taken. To obtain the maximum value of $\vec{\pi _{N}}$, one
only needs to transfer the momentum ``$(N-1)|\Lambda |$" of $N^{th}$
particle from with respect to particle 1 to with respect to the (N-1)-body's
c.m. frame, which can be simply obtained through Eq. (\ref{jaco}):%
\begin{eqnarray}
\vec{\pi}_{N-1} &=&\frac{N-1}{N}\left( \vec{p_{N}}-\frac{1}{N-1}(\vec{p_{1}}+%
\vec{p_{2}}+...+\vec{p}_{N-1})\right) ,  \notag \\
\vec{\pi}_{N-1}^{\max } &=&\frac{N-1}{N}\left( N-1-\frac{1}{N-1}%
(0+1+2+...+N-2)\right) \vec{q}  \notag \\
&=&\frac{N-1}{2}\vec{q}.  \label{res}
\end{eqnarray}%
Thus, the maximum value of the $(N-1)^{th}$ Jacobi momentum $|\vec{\pi}%
_{N-1}^{\max }|=(N-1)\Lambda $.

In ab-initio calculations, the total model space where a N-body system
occupied corresponds exactly to the space spanned by the wavefunctions in
the Jacobi coordinate representation. For example, when the wavefunctions
are expanded in Harmonic oscillator (HO) basis, the maximum allowed number
of HO excitations for an A-particle system $N_{A}$ can be defined as\cite%
{bruce}:%
\begin{equation}
N_{A}=\sum\limits_{i=1}^{A-1}(2n_{\pi _{i}}+l_{\pi _{i}}),  \label{nc}
\end{equation}%
where $n_{\pi _{i}},$ $l_{\pi _{i}}$ are the principle and angular momentum
quantum number associated with Jacobi coordinate $\pi _{i}$. Note that the
link between $n_{\pi _{i}},$ $l_{\pi _{i}}$ and the corresponding momentum
cutoff of each Jacobi coordinate, $|\vec{\pi}_{i}|$ is provided by $|\vec{\pi%
}_{i}|=\sqrt{M(2n_{_{\pi _{i}}}+l_{\pi _{i}}+7/2)\hbar \omega }$\cite{extra}%
, with $M$ the nucleon mass, $\hbar $ the Planck constant and $\omega $ the
oscillator strength. The total maximum model space for an A-particle ($A>1$)
system is then%
\begin{eqnarray}
N_{A}^{\max } &=&\sum\limits_{i=1}^{A-1}(2n_{\pi _{i}}^{\max }+l_{\pi
_{i}}+7/2)  \notag \\
&=&\frac{\Lambda ^{2}}{M\hbar \omega }\sum\limits_{i=1}^{A-1}i^{2}  \notag \\
&=&\frac{A(A-1)(2A-1)}{6}\frac{\Lambda ^{2}}{M\hbar \omega } \notag \\
&=&\frac{A(A-1)(2A-1)}{6}N^{\max}_2.  \label{total}
\end{eqnarray}

\bigskip

\section{Discussion}

\bigskip First, we note that Eq. (\ref{total}) denotes the exact model space
to be used in principle in the ab-initio calculations. In other words, if
the inter-particle interaction is established or renormalized within a
finite model space (with $\Lambda $ and $\omega $ specified), then, instead
of extrapolating to infinity, one should stop at the finite size of model
space specified in Eq. (\ref{total}) in an $A$-body calculation.
Additionally, even in cases where one suppresses the interaction (normally
by exponential regulators) after the momentum $k>\Lambda $, so that the
interaction still occupies the entire space, adopting a model space higher
than the one defined in Eq. (\ref{total}) will just capture the artifact of
unimportant physics in the high momentum tail of the interaction. We note
that, in usual ab-initio calculations, the contribution given by including
model spaces higher than $N_{A}^{\max }$ given in Eq. (\ref{total}) does not
vanish automatically. The kernel of the interaction $V_{ij}$ restricts the
interaction between particle $i$ and $j$ up to the shells correspond to the
maximum momentum $\Lambda ,$ but the rest of the configurations could go up
to arbitrary high $n_{\pi _{i}}$, $l_{\pi _{i}}$ and still give contribution
to the matrix element. Thus, in calculations where $N_{A}>$ $N_{A}^{\max }$
is practical, one should consider the results obtained up to $N_{A}=N_{A}^{\max
}$ only.

From the results of Eq. (\ref{total}), one immediately sees that the
HO-basis model space grows as cubic power in number of particles, i.e., $%
A^{3}$. Translating the space into the total accumulated momentum $%
P_{A}^{\max }$,%
\begin{equation}
P_{A}^{\max }\equiv \sum\limits_{i=1}^{A}(i-1)\Lambda =\frac{A(A-1)}{2}%
\Lambda ,  \label{pa}
\end{equation}%
one still has an $A^{2}$ grows in the maximum momentum an $A$-body system
could have. Thus, even starting with a small two-body cutoff $\Lambda $, the
model space where the $A$-body system lives is enriched quickly by the
number of particles. 

If one is only interested in probing physical phenomena
up to a fixed energy scale for a wide range of particle number $A$, then the
required two-body model space would decrease drastically
with the number of particle of the system. For example, suppose one starts
with a two-body interaction $V_{2}(p,p^{\prime })$ with flexible form and
parameters, where $p^{(\prime )}$ is the incoming (outgoing) momentum in the
c.m. frame. Then to encode physics up to the same energy scale into the
parameters of $V_{2}(p,p^{\prime })$, the minimum required momentum space
would range from $p^{(\prime )}\in[0,\Lambda ]$ for a 2-particle system to $%
p^{(\prime )}\in [0,\frac{\Lambda }{A^{2}}]$ for an $A$-particle system.
Whether there exists an EFT expansion on the parametrization of $%
V_{2}(p,p^{\prime })$ is another question.

In other words, if there exists a way to renormalize the interaction under
an EFT to arbitrary low cutoffs $\Lambda $, then the required (minimum)
two-body cutoff to describe a physical phenomenon (up to a given energy
scale) in many-body systems would be much smaller than the required cutoff
to describe the few-body systems. In nuclear physics, this required cutoff
quickly goes below the pion-exchange threshold. Take the case of pure
neutron matter as an example. In this extreme case (where $A\rightarrow
\infty $), it was demonstrated that the equation of state (EOS) up to twice
of the saturation density is not far from a simple parametrization governed
by the Bertch parameter\cite{Ber00,bira-uni}. It also suggests that the
effect of pion-exchange between nucleons can be integrated out and
re-parametrized into an EFT-like expansion around the unitarity limit\cite%
{denis,denis2,denis3}, if one just wants to obtain a reasonable EOS up to
twice of the saturation density. Moreover, by considering beyond mean field
effects, it could be possible to establish an EFT-based, Skyrme-like
interaction, to describe general properties for systems consist of many
protons and neutrons\cite{yang-edf,more-edf}.

Another case of interested is the recent attempts to apply an
unitarity-based pionless EFT into the calculations of light nuclei properties%
\cite{bira-uni2}, where, at leading order (LO), one first adjusts the
interaction at three-body level instead of fitting to two-body observables%
\footnote{%
The two-body interaction is kept at unitairty limit at LO.}. Based on the power
counting analysis, at least in the fermionic case, the details of the
two-body nucleon-nucleon interaction (i.e., corrections on top of unitarity)
only starts to contribute at NLO. As one can see in Eq. (\ref{pa}), only $1/3$ of $\Lambda$ in the interaction is needed to
describe a three-body observable than to describe a two-body observable at the same
energy scale\footnote{This also suggests that the minimum cutoff for a three-body force is $1/3$ of the two-body cutoff $\Lambda$, provided that the renromalization is performed consistently and 
one does not probe physics of the three-body system higher than $k\ge\Lambda/3$.}. Therefore, on top of the advantages analyzed in the power counting, an earlier conversion of the three-body observable into
the low energy constants in the theory could capture important physics into
the interaction within a smaller model space and potentially
fasten the convergence for high-body calculations.

\section{Conclusion}

In this work, a direct link between two- and A-body model space is derived.
It is found that instead of infinity, one needs to extrapolate to a finite
many-body model space in ab-initio calculations when interactions
renormalized in a given model space are adopted. In addition, the results
suggest that to describe a system equally well up to a fixed energy scale,
it is possible to \textquotedblleft integrated out" more ultra-violate
physics in the many-body cases, provided that the errors is quantified by
the EFT principle.

\bigskip

\begin{acknowledgments}
The authors thanks G. Hupin and H. W. Griesshammer for useful discussion at different stage of the work.
This project has received funding from the European Unions Horizon 2020
research and innovation program under grant agreement No. 654002 and US Department of Energy under contract DE-SC0015393.
\end{acknowledgments}

\end{document}